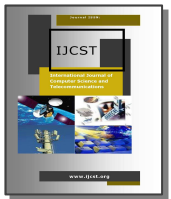

# Network Selection Decision Based on Handover History in Heterogeneous Wireless Networks

Lahby Mohamed, Cherkaoui Leghris and Adib Abdellah

Department of Computer Science, LIM Lab, Faculty of Sciences and Technology of Mohammedia,

B.P. 146, Mohammedia, Morocco

{mlahby, cleghris, adib_adbe}@yahoo.fr

*Abstract*— In recent years, the mobile devices are equipped with several wireless interfaces in heterogeneous environments which integrate a multitude of radio access technologies (RAT's). The evolution of these technologies will allow the users to benefit simultaneously from these RAT's. However, the most important issue is how to choose the most suitable access network for mobile's user which can be used as long as possible for communication. To achieve this issue, this paper proposes a new approach for network selection decision based on Saaty's Fuzzy Analytical Hierarchy Process (FAHP) and the Technique for Order Preference by Similarity to an Ideal Solution (TOPSIS). The FAHP method is used to determine a weight for each criterion, and the TOPSIS method is applied to rank the alternatives. Simulation results are presented to illustrate the effectiveness of our new approach for network selection.

## I. INTRODUCTION

IN recent years, the mobile devices are equipped with several wireless interfaces in heterogeneous environments which integrate a multitude of radio access technologies (RAT's) such as wireless technologies (802.11a, 802.11b, 802.15, 802.16, etc.) and cellular networks (GPRS, UMTS, HSDPA, LTE, etc.). The evolution of these technologies will allow the users to benefit simultaneously from these RAT's and they can also use various services offered by each type of access network.

The most important issue in RAT's, is to provide ubiquitous access for the end users, under the principle "Always Best Connected" (ABC) [1], to achieve this issue a vertical handoff decision [2] is intended to determine whether a vertical handoff should be initiated, and to choose the most suitable network in terms of quality of service (QoS) for mobile users. The vertical handover process can be divided into three steps:

1) Handover initiation: it contains some preparation for handoff such as the measurement of received signal strength (RSS), QoS, security, battery level, etc.

2) Network selection : it consists on choosing the most

suitable network access among those available to perform a handover.

3) Handover execution: it consists on establishing the target access network by using mobile IP protocol (MIP).

The network selection problem is the most important key of the vertical handover decision. For that our work focuses on the optimization of the network selection decision for users in order to support many services with best QoS and let the users stay in current access network as long as possible.

However, no single wireless network technology is considered to be more favorable than other technologies in terms of QoS. In other words, each network access in RAT's seems to be specifically characterized by the bandwidth offered, the coverage ensured by the network as well as the cost to deliver the service. Moreover, there is some kind of complementarity between these various networks; for example, 801.11a offers a higher bandwidth with a cover limited, while UMTS ensures a large cover with lower bandwidth. The network selection algorithm depends on multiple criteria which are:

- From terminal side: battery, velocity, etc.

- From service side: QoS level, security level, etc.

- From network side: provider's profile, current QoS parameters, etc.

- From user side: users preferences, perceived QoS, etc.

In the other hand the network selection problem can be tackled with several schemes and decision algorithms such as multi-attribute decision making (MADM) methods, genetic algorithms, and fuzzy logic. In [3], the authors have proposed a new strategy to solve the vertical handover decision problem by using the fuzzy multi-attribute decision making methods (Fuzzy MADM). In [4], the authors have proposed also an intelligent approach for vertical handover using fuzzy logic. In [5] and [6], the network selection algorithm is based on AHP and gray relation analysis (GRA) two MADM methods. The AHP method is used to weigh each criterion and GRA method is applied to rank the alternatives. In [7] and [8], the network selection algorithm combines two MADM methods AHP and TOPSIS. The AHP method is used to get weights of the criteria and TOPSIS method is applied to determine the ranking of access network.

Due to great number of criteria and algorithms which can be used in network selection, the most challenging problems focus in selecting the appropriate criteria and definition of a strategy which can exploit these criteria. According to nature of network selection problem MADM algorithms, represent a promising solution to select the most suitable network in terms of quality of service (QoS) for mobile users.

However the major limitation of MADM methods is the ping pong effect provided in the network selection decision. The ping pong effect occurs when the terminal mobile performs excessive handoffs for a given time which causing the higher number of handoffs. This phenomenon can led to increasing in power consumption and the decreasing in throughput.

In order to deal with this problem, we propose an intelligent network selection approach based on new history attribute. Our new approach is based on Saaty's Fuzzy Analytical Hierarchy Process (FAHP) and the Technique for Order Preference by Similarity to an Ideal Solution (TOPSIS). The FAHP method is used to determine a weight for each criterion, and the TOPSIS method is applied to rank the alternatives. The history criterion is introduced to reduce the number of handoff and ensure that the terminal mobile stay in current access network as long as possible.

This paper is organized as follows. Section II describes Multi Attribute Decision Making methods (MADM). Section III presents our network selection algorithm based on FAHP and TOPSIS two MADM methods. Section IV includes the simulations and results. Section V concludes this paper.

## II. MULTI ATTRIBUTE DECISION MAKING METHODS

### A. FAHP

The Fuzzy Analytic Hierarchy Process (FAHP) is one of the extensive multi-attribute decision making. Fuzzy AHP is an extension of Analytic Hierarchy Process (AHP) [10] has been developed to solve hierarchical fuzzy problems [11].

In the fuzzy AHP procedure, the pair-wise comparisons in the judgment matrix are fuzzy numbers that are modified according to the designers focus.

The FAHP is based on four stages:

1) Construct of the structuring hierarchy: A problem is decomposed into a hierarchy, this one contains three levels: the overall objective is placed at the topmost level of the hierarchy, the subsequent levels presents the decision factors and the alternative solution are located at the bottom level.

2) Construct of the pair-wise comparisons: to establish a decision, FAHP builds the pair-wise matrix comparison such as:

$$A = \begin{bmatrix} r_{11} & r_{12} & \cdots & \cdots & r_{1n} \\ r_{21} & r_{22} & \cdots & \cdots & r_{2n} \\ \vdots & \vdots & \vdots & \vdots & \vdots \\ r_{n1} & \cdots & \cdots & \cdots & r_{nn} \end{bmatrix} \begin{cases} r_{ii} = 0.5 \\ r_{ij} + r_{ji} = 1 \end{cases} \quad (1)$$

Elements $r_{ij}$ are obtained from the table I, it contains the preference scales.

### TABLE I
SAATY'S SCALE FOR FUZZY PAIR-WISE COMPARISON

| Saaty's scale | The relative importance of the two sub-elements |
|---|---|
| 0.5 | Equally important |
| 0.55(OR 0.5 0.6) | Slighly important |
| 0.65(OR 0.6 0.7) | Important |
| 0.75(OR 0.7 0.8) | Strongly important |
| 0.85(OR 0.8 0.9) | Very strongly important |
| 0.95(OR 0.9 1.0) | Extremely important |

3) Calculating the weights of criterion: the weights of the decision factor i can be calculated by:

$$W_i = \frac{b_i}{\sum_{j=1}^{n} b_i} \quad and \quad \sum_{i=1}^{n} W_i = 1 \quad (2)$$

$$Where \quad b_i = \frac{1}{\left[\sum_{j=1}^{n} \frac{1}{r_{ij}}\right] - n} \quad (3)$$

4) Calculating the coherence ratio (CR): to test consistency of a pairwise comparison, a consistency ratio (CR) can be calculated as

$$CR = \frac{CI}{RI} \quad (4)$$

Where the consistency index (CI) can be calculated by:

$$CI = \frac{\left|\sum_{i=1}^{n} \frac{(AW)_i}{nW_i}\right|}{n-1} \quad (5)$$

RI is the index of matrix coherence, the various values of RI are shown in table II.
If the CR is less than 0.1, the pair-wise comparison is considered acceptable.

### TABLE II
VALUE OF RANDOM CONSISTENCY INDEX RI

| CRITERIA | 3 | 4 | 5 | 6 | 7 | 8 | 9 | 10 |
|---|---|---|---|---|---|---|---|---|
| RI | 0.58 | 0.90 | 1.12 | 1.24 | 1.32 | 1.41 | 1.45 | 1.49 |

### B. TOPSIS

Technique for order preferences by similarity to an ideal solution (TOPSIS), known as a classical multiple attribute decision-making (MADM) method, has been developed in 1981 [12]. The basic principle of the TOPSIS is that the chosen alternative should have the shortest distance from the positive ideal solution and the farthest distance from the negative ideal solution. The procedure can be categorized in six steps:

1) Construct of the decision matrix: the decision matrix is expressed as

$$D = \begin{bmatrix} d_{11} & d_{12} & \cdots & \cdots & d_{1m} \\ d_{21} & d_{22} & \ddots & & d_{2m} \\ \vdots & \vdots & \ddots & \ddots & \vdots \\ d_{n1} & d_{n2} & \cdots & \cdots & d_{nm} \end{bmatrix} \quad (6)$$

Where $d_{ij}$ is the rating of the alternative $A_i$ with respect to the criterion $C_j$

2) Construct the normalized decision matrix: each element $r_{ij}$ is obtained by the euclidean normalization;

$$r_{ij} = \frac{d_{ij}}{\sqrt{\sum_{i=1}^{m} d_{ij}^2}}, \; i=1,\ldots,m \text{ and } j=1,\ldots,n. \quad (7)$$

3) Construct the weighted normalized decision matrix: The weighted normalized decision matrix $v_{ij}$ is computed as:

$$v_{ij} = W_i * r_{ij} \; where \; \sum_{j=1}^{n} W_i = 1 \quad (8)$$

4) Determination of the ideal solution $A^*$ and the anti-ideal solution $A^-$:

$$A^* = [V_1^*, \ldots, V_m^*] \; and \; A^- = [V_1^-, \ldots, V_m^-] \quad (9)$$

• For desirable criteria:

$$V_i^* = \max\{v_{ij}, j = 1, \ldots, n\} \quad (10)$$

$$V_i^- = \min\{v_{ij}, j = 1, \ldots, n\} \quad (11)$$

• For undesirable criteria:

$$V_i^* = \min\{v_{ij}, j = 1, \ldots, n\} \quad (12)$$

$$V_i^- = \max\{v_{ij}, j = 1, \ldots, n\} \quad (13)$$

5) Calculation of the similarity distance:

$$S_i^* = \sqrt{\sum_{j=1}^{m}\left(V_i^* - v_{ij}\right)^2}, j = 1, \ldots, n \quad (14)$$

And

$$S_i^- = \sqrt{\sum_{j=1}^{m}\left(V_i^- - v_{ij}\right)^2}, j = 1, \ldots, n \quad (15)$$

6) Ranking:

$$C_j^* = \frac{S_i^-}{S_i^* + S_i^-}, j = 1, \ldots, n \quad (16)$$

A set of alternatives can be ranked according to the decreasing order of $C_j^*$.

### III.   ACCESS NETWORK SELECTION ALGORITHM

In order to reduce the number of handoffs and to ensure that the mobile terminal can stay in the current access networks as long as possible, we present our new intelligent network selection approach based on two MADM methods such as Fuzzy AHP method and TOPSIS method. The new approach introduces a new criterion namely history. This attribute allows to memorize the overall score given to the available alternative by using the TOPSIS method (history value is $C_j^*$).

The algorithm assumes wireless overlay networks which entail three heterogeneous networks such as UMTS, WLAN and WIMAX. Instead of using six attributes associated in each access network which are: Cost per Byte (CB), Available Bandwidth (AB), Security (S), Packet Delay (D), Packet Jitter (J) and Packet Loss (L), we add a new History criterion (H).

Fig. 1 exhibits the three levels Fuzzy AHP hierarchy for the classical network selection algorithm which don't taking into consideration the history attribute. The level 1 includes three criteria QoS, security and cost, the level 2 includes four QoS parameters such as AB, D, J and L and the level 3 includes three available networks UTMS, WIFI and WIMAX.

In the other hand Fig. 2 exhibits the three levels Fuzzy AHP hierarchy for our new network selection approach which takes into consideration the history attribute. The level 1 includes four criteria QoS, security, cost and history, the level 2 and level 3 are the same of level 2 and level 3 of Fig. 1 respectively.

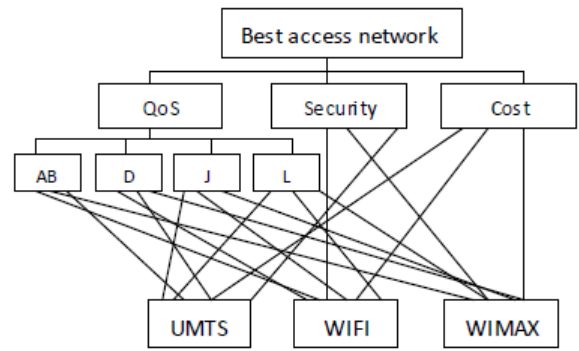

Fig. 1. FAHP hierarchy for classical network selection decision

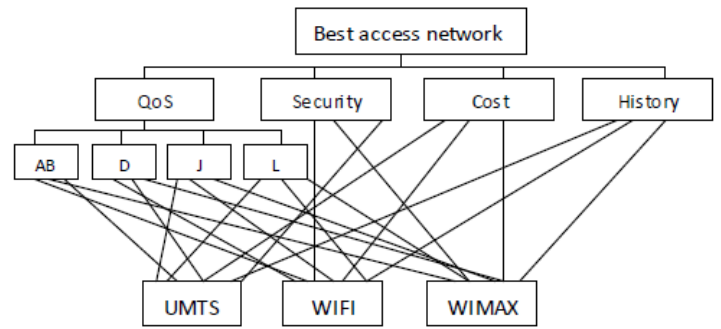

Fig. 2. FAHP hierarchy for our network selection decision

Based on the specific characteristics of the traffic type [13], our new approach can be categorized in four steps:

1) Assign weights to level-1-criteria: the FAHP method is used to get weights of the criteria of level 1.

2) Assign weights to level-2-criteria: the FAHP method is used to get weights of the criteria of level 2.

3) Assign weights to level-3-alternatives: the weight vector of each available network is calculated by multiplication of the weight vector obtained in level 1 with the weight vector obtained in level 2.

4) Select the best access network: the method TOPSIS is applied to rank the available networks and select the

access network that has the highest value of $C_j^*$ (see the steps of TOPSIS method).

## IV. SIMULATION AND RESULTS

In order to illustrate the effectiveness of our new approach based on Fuzzy AHP and TOPSIS which takes into consideration a new history attribute we present performance comparison between two algorithms which are:

- TOPSIS-1: it's the classical network selection algorithm based on FAHP and TOPSIS which don't considering the history attribute.
- TOPSIS-2: it's our new network selection approach based on FAHP and TOPSIS including the history attribute.

The performance evaluation for four traffic classes [13] namely background, conversational, interactive and streaming is focused on reducing the number of handoffs. In each simulation the two algorithms were run in 10 vertical handoff decision points. For TOPSIS-1 method the history criterion for each access network has no effect on our simulation.

During the simulation, for each candidate networks, the measures of six attributes CB, AB, S, D, J and L are randomly varied according to the ranges shown in table III. Furthermore the value of history criterion is initialized by 1, after the value of $H_{i+1}$ is equal to $C_j^*$ in iteration i+1 where $C_j^*$ is the score of TOPSIS method obtained in iteration i.

TABLE III
ATTRIBUTE VALUES FOR THE CANDIDATE NETWORKS

| Criteria Network | CB (%) | S (%) | AB (mbps) | D (ms) | J (ms) | L (per10⁶) | H (%) |
|---|---|---|---|---|---|---|---|
| UMTS | 60 | 70 | 0.1-2 | 25-50 | 5-10 | 20-80 | 100 |
| WLAN | 10 | 50 | 1-11 | 100-150 | 10-20 | 20-80 | 100 |
| WIMAX | 40 | 60 | 1-60 | 60-100 | 3-10 | 20-80 | 100 |

### A. The Simulation 1

In this simulation, the background traffic is analyzed, we present the performance comparison between two vertical handoff algorithms such as TOPSIS-1 and TOPSIS-2.

A set of importance weights of the criteria of TOPSIS-1 and TOPSIS-2 are displayed in Fig. 3.

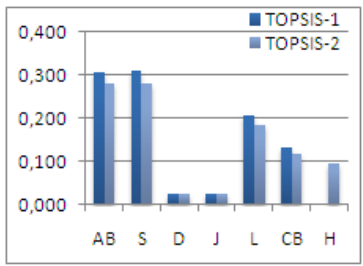

Fig. 3. Weights associated with the criteria for background traffic

Fig. 4 shows the average value of the number of handoffs in 10 vertical handoff decision points. We notice that, the TOPSIS-1 method reduces the number of handoffs with a value of 50%, and the TOPSIS-2 method reduces the number of handoffs with a value of 30%.

So for background traffic, our method based on TOPSIS-2 can reduce the number of handoffs better than TOPSIS-1.

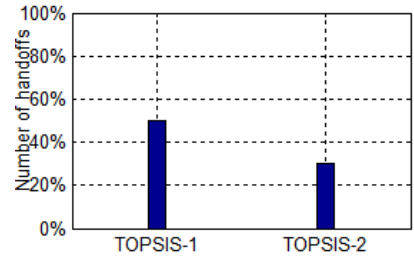

Fig. 4. Average of number of handoffs for background traffic

### B. The simulation 2

In this simulation, the traffic analyzed is conversational traffic; we present the performance comparison between two vertical handoff algorithms such as TOPSIS-1 and TOPSIS-2.

A set of importance weights of the criteria of TOPSIS-1 and TOPSIS-2 are displayed in Fig. 5.

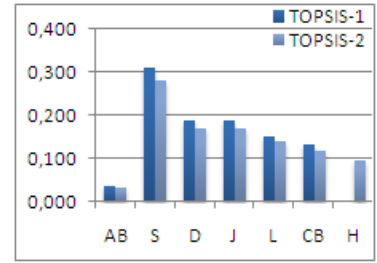

Fig. 5. Weights associated with the criteria for conversational traffic

Fig. 6 shows the average value of the number of handoffs in 10 vertical handoff decision points. We notice that, the TOPSIS-1 method reduces the number of handoffs with a value of 60%, and the TOPSIS-2 method reduces the number of handoffs with a value of 40%.

So for conversational traffic, our method based on TOPSIS-2 can reduce the number of handoffs better than TOPSIS-1.

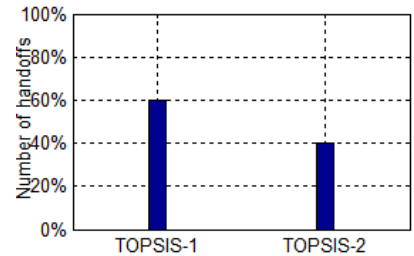

Fig. 6. Average of number of handoffs for conversational traffic

### C. The simulation 3

In this simulation, the traffic analyzed is interactive traffic, we present the performance comparison between two vertical handoff algorithms such as TOPSIS-1 and TOPSIS-2.

A set of importance weights of the criteria of TOPSIS-1 and TOPSIS-2 are displayed in Fig. 7.

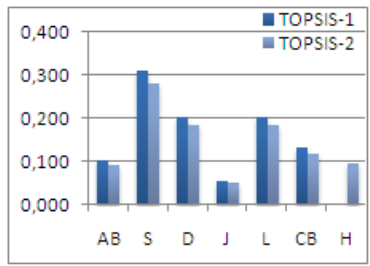

Fig. 7. Weights associated with the criteria for interactive traffic

Fig. 8 shows the average value of the number of handoffs in 10 vertical handoff decision points. We notice that, the TOPSIS-1 method reduces the number of handoffs with a value of 70%, and the TOPSIS-2 method reduces the number of handoffs with a value of 40%.

So for interactive traffic, our method based on TOPSIS-2 can reduce the number of handoffs better than TOPSIS-1.

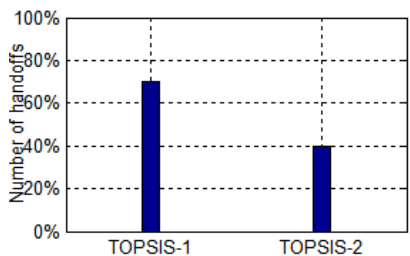

Fig. 8. Average of number of handoffs for interactive traffic

### D. The simulation 4

In this simulation, the traffic analyzed is streaming traffic, we present the performance comparison between two vertical handoff algorithms such as TOPSIS-1 and TOPSIS-2.

A set of importance weights of the criteria of TOPSIS-1 and TOPSIS-2 are displayed in Fig. 9.

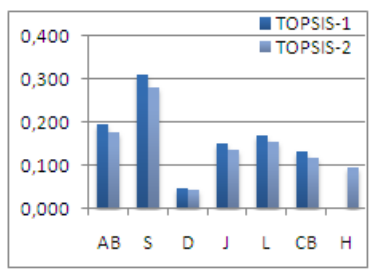

Fig. 9. Weights associated with the criteria for streaming traffic

Fig. 10 shows the average value of the number of handoffs in 10 vertical handoff decision points. We notice that, the TOPSIS-1 method reduces the number of handoffs with a value of 40%, and the TOPSIS-2 method reduces the number of handoffs with a value of 20%.

So for streaming traffic, our method based on TOPSIS-2 can reduce the number of handoffs better than TOPSIS-1.

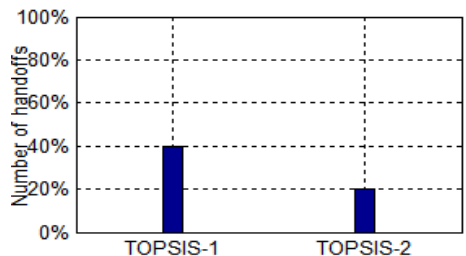

Fig. 10. Average of number of handoffs for streaming traffic

### V. Conclusion

In this work, we have proposed a new approach for network selection based on Fuzzy AHP method and TOPSIS method. The proposed approach takes into consideration a new criterion namely history. The use of this attribute helps to deal with the ping-pong effect by reducing the number of handoffs, and keeping the terminal mobile as long as possible in the current access network.

The simulation results show that, our method based on TOPSIS-2 method provide best performance concerning the number of handoffs than TOPSIS-1 for all four traffic classes namely, background, conversational, interactive and streaming.

### References

[1]  E. Gustafsson and A. Jonsson, "Always best connected", IEEE Wireless Communications Magazine, vol.10, no.1,pp.49-55, Feb. 2003.

[2]  H. Wang, R. Katz, J. Giese "Policy-enabled handoffs across heterogeneous wireless networks", Second IEEE Worshop on Mobile Computing systems and Applications, WMCSA. pp. 51-60, February 1999.

[3]  W. Zhang "Handover Decision Using Fuzzy MADM in Heterogeneous Networks", Wireless Communications and Networking Conference, WCNC. 2004 IEEE, Vol. 2, pp.653-658, March 2004.

[4]  H. Attaullah and al,"Intelligent vertical handover decision model to improve QoS", In Proceedings of ICDIM, pp.119-124, 2008.

[5]  J. Fu, J. Wu, J. Zhang, L. Ping, and Z. Li, " Novel AHP and GRA Based Handover Decision Mechanism in Heterogeneous Wireless Networks", in Proc. CICLing (2), pp. 213-220, 2010.

[6]  Wang Yafang; Cui Huimin; Zhang Jinyan. "Network access selection algorithm based on the Analytic Hierarehy Process and Gray Relation Analysis", in Proc. New Trends in Information Science and Service Science NISS'2010, pp. 503-506, May 2010.

[7]  M. Lahby, C. Leghris. and A. Adib. "A Hybrid Approach for Network Selection in Heterogeneous Multi-Access Environments", In the Proceedings of the 4th IFIP International Conference on New Technologies, Mobility and Security (NTMS), pp. 1-5, Feb 2011.

[8]  F. Bari and V. Leung, "Multi-attribute network selection by iterative TOPSIS for heterogeneous wireless access", 4[th] IEEE Consumer Communications and Networking Conference, pp. 808-812, January 2007.

[9]  E. Dimitris, and al. "Application of Fuzzy AHP and ELECTRE to Network Selection", In the Mobile Lightweight Wireless Systems Volume 13, Pages 63-73, 2009.

[10] T. L. Saaty "Decision Making for Leaders: The Analytic hierarchy Process for Decisions in a Complex World", RWS Publications, 1988.

[11] Mahmoodzadeh, S. and al "Project Selection by Using Fuzzy AHP and TOPSIS Technique", In proceeding of World Academy of Science, Engineering and Technology, October 2007, vol. 24, (2007).

[12] E. Triantaphyllou "Multi-Criteria Decision Making Methods: A Comparative Study", Kluwer academic publishers, Applied optimization series, Vol. 44, 2002.

[13] "3GPP, QoS Concepts and Architecture" 2005, tS 22.107 (v 6.3.0).